\documentclass[a4paper]{jpconf}

\usepackage{amssymb}
\usepackage{amsfonts}
\usepackage{amsmath}
\usepackage{bm}
\usepackage{epsfig}
\usepackage{epsf}
\usepackage[]{graphicx}

\usepackage{color}
\usepackage{cite}
\usepackage{xspace}

\renewcommand{\d}{{\rm d}}

\newcommand {\calA} {{\cal A}}

\newcommand {\E}  {{\varepsilon}}
\newcommand {\om}  {{\omega}}
\newcommand {\Om}  {{\Omega}}

\newcommand {\Ld}  {L_{\rm d}}

\newcommand{\Nacc}{{N_{\rm acc}}}
\newcommand{\Lch}{{L_{\rm ch}}}

\newcommand{\lamu}{{\lambda_{\rm u}}}

\begin{document}


\title
{Sub-GeV Electron and Positron Channeling in Straight, Bent and Periodically Bent 
Silicon Crystals}

\author{G B Sushko$\P$, 
A V Korol$\P\S$,
Walter Greiner$\P$,
A V Solov'yov$\S$
\footnote[1]{On leave from A.F. Ioffe Physical-Technical Institute, St. Petersburg, Russia}}

\address{$\P$ Frankfurt Institute for Advanced Studies, 
Ruth-Moufang-Str. 1, 60438 Frankfurt am Main, Germany}
\address{$\S$ St. Petersburg State Maritime University, 
Leninsky ave. 101, 198262 St. Petersburg, Russia}

\ead{korol@fias.uni-frankfurt.de}

\begin{abstract}
Preliminary results of numerical simulations of electron and positron channeling
and emission spectra are reported for straight, uniformly bent and periodically
bent  silicon crystal. 
The projectile trajectories are computed using the newly developed module
\cite{NewPaper_2013} of the MBN  Explorer package 
\cite{MBN_ExplorerPaper,MBN_ExplorerSite}.
The electron and positron channeling along Si(110) and Si(111) 
crystallographic planes are studied for the 
projectile energies 195--855 MeV.
\end{abstract}

\section{Introduction}

The basic effect of the channeling process in a crystal 
is in an anomalously large distance which particle can penetrate moving along a 
crystallographic plane (the planar channeling) or an axis (the axial channeling) 
and experiencing the collective action of the electrostatic field of the 
lattice ions \cite{Lindhard}.
The field is repulsive at small distances for positively charged 
particles and, therefore, they are steered into the inter-atomic region,
while negatively charged projectiles move in the close vicinity of ion
strings or planes.\footnote{Channeling effect can be discussed not 
only for crystals but, in principle, for any structured material which provides "passages",
moving along which a projectile has much lower value of the mean square 
of the multiple scattering angle than when moving along any random direction.
The examples of such materials are nanotubes and fullerites, for which the channeling
effects has been also investigated, see, e.g., Ref. \cite{ArtruEtAl_PhysRep2005}.}

Channeling of charged particles is accompanied by the channeling radiation 
\cite{ChRad:Kumakhov1976} which arises due to the
transverse motion (the channeling oscillations) of the particle inside 
the channel under the action of the interplanar or axial field.
The intensity of the radiation depends on the type of the projectile,
on its energy, and on the type of crystal and crystallographic plane (axis).
In a bent crystal, the synchrotron radiation
appears as a result of circular motion of the channeling particle along the 
bent crystallographic planes. 
The condition of stable channeling in bent crystals 
\cite{Tsyganov1976,BiryukovChesnokovKotovBook} inplies
the bending radius $R$ to be much smaller than the (typical) curvature radius of 
the channeling oscillations. 
Therefore, the synchrotron radiation modifies the soft-photon part of the 
emission spectrum.
In a periodically bent crystal, additional mechanism of radiation appears
to the undulating motion of channeling particles which follow the periodic 
bending of crystallographic planes.
The concept of a crystalline undulator (CU) was formulated quite recently  
\cite{KSG1998,KSG_review_1999}.
By means of CU it is feasible to produce monochromatic undulator-like radiation 
in the hundreds of keV up to the MeV photon energy range. 
The intensity and characteristic frequencies of the radiation can 
be varied by changing the type of channeling particles,
the beam energy, the crystal type  and the parameters of periodic bending.
(see recent review \cite{ChannelingBook2013}  for more details).

In recent years, several experiments were carried out to measure the 
channeling parameters and the characteristics of emitted radiation 
of sub-GeV light projectiles.
These include the attempts made  \cite{BaranovEtAl_CU_2006} 
or planned to be made \cite{Backe_EtAl_2011a} 
to detect the radiation from a positron-based CU.
More recently, a series of the experiments with straight, bent  and 
periodically bent crystals
have been carried out with 195--855 MeV electron beams  
at the Mainz Microtron (Germany) facility
\cite{Backe_EtAl_2010,Backe_EtAl_2011}.
The CUs, used in the experiment, 
were manufactured in Aarhus University (Denmark) using the molecular beam
epitaxy technology to produce strained-layer Si$_{1-x}$Ge$_{x}$ 
superlattices with varying germanium content as described in 
\cite{MikkelsenUggerhoj2000,Darmstadt01}.
Another set of experiments with diamond CUs is scheduled for the year 
2013 at the SLAC facility (USA) with 10\dots 20 GeV electron 
beam \cite{Uggerhoj_2012}.

Theoretical support of the ongoing and future experiments as well as accumulation
of numerical data on channeling and radiative processes of ultra-relativistic 
projectiles in crystals of various content and structure must be based on an 
accurate procedure which allows one to simulate the trajectories 
corresponding to the channeling and non-channeling regimes.
Recently, a universal code to simulate trajectories
of various projectiles (positively and negatively charged, light and heavy) 
in an arbitrary scattering medium, either structured (straight, bent and
periodically crystals, superlattices, nanotubes etc) or amorphous (solids, 
liquids) has been developed as a new module
\cite{NewPaper_2013} of the MBN  Explorer package 
\cite{MBN_ExplorerPaper,MBN_ExplorerSite}.
To simulate propagation of particles through media the algorithms used in modern 
molecular dynamics (MD) codes were applied.
Verification of the code against available experimental data
as well as against predictions of other theoretical models 
were carried out for $6.7$ GeV and $855$ MeV electrons and positrons in Si(110) 
as well as in amorphous Si \cite{NewPaper_2013}.
In the cited paper, critical analysis was carried out of the underlying physical model 
and the algorithm implemented in the recent code for electron channeling described in 
Refs. \cite{KKSG_simulation_straight}.
It was shown, that the specific model for electron--atom scattering
leads in a noticeable overestimation of the mean scattering angle.
As a results, the data on the dechanneling lengths presented in  
\cite{KKSG_simulation_straight} are incorrect.

In the present paper we present new results on electron and positron channeling
and emission spectra in straight, uniformly bent and periodically
bent silicon crystal. 
The electron and positron channeling along Si(110) and Si(111) 
crystallographic planes are studied for the 
projectile energies 195--855 MeV.

\section{Channeling Radiation of Electrons in  Si (110)}

\begin{figure} [h]
\centering
\includegraphics[scale=0.31,clip]{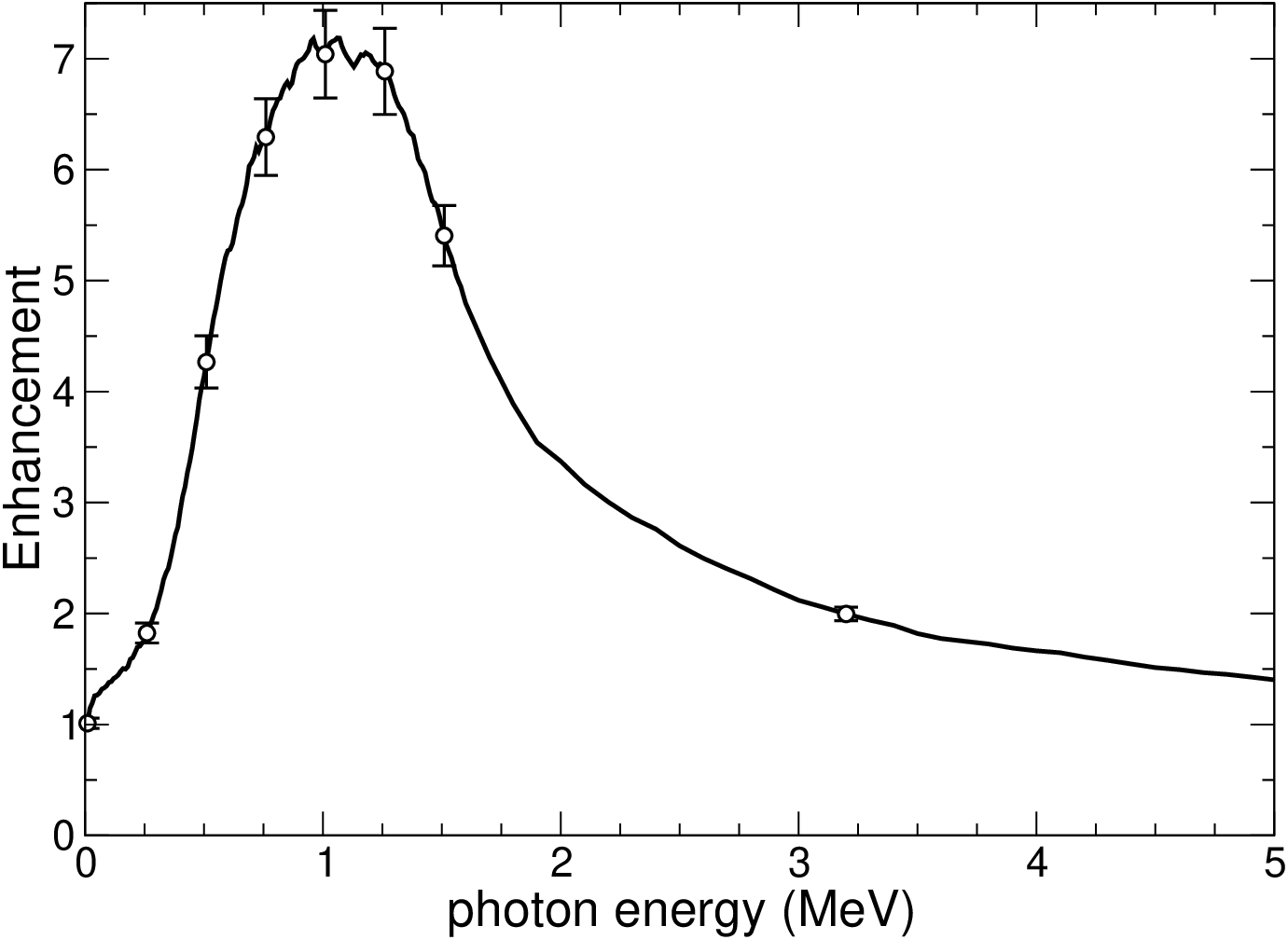}
\hspace*{0.5cm}
\includegraphics[scale=0.31,clip]{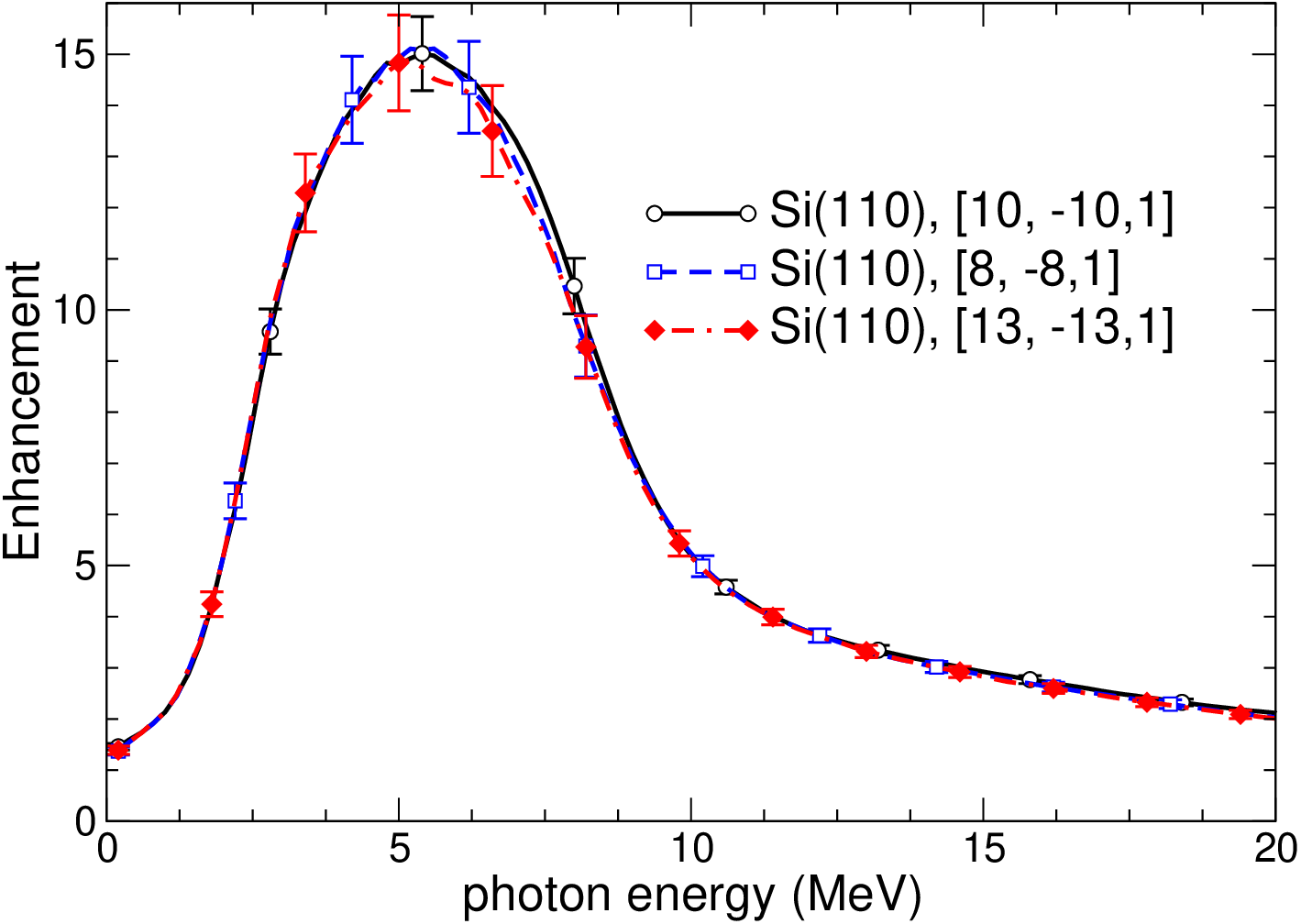}
\caption{
Enhancement factor
for 270 MeV (left panel) and for 855 MeV (right panel) 
electron channeling along Si(110) plane in $L=50$ $\mu$m thick straight crystal.
The 270 MeV data refer to the maximum emission angle 
$\theta_{\max}=0.24$ mrad, the  855 MeV data -- to $\theta_{\max}=0.21$ mrad.
Three curves on the right panel correspond to three different directions of the incident
beam, as indicated.
}
\label{Figures_01_02.fig}
\end{figure}

A relevant benchmark for our simulations are the channeling spectra
of 195 \dots 855 MeV electrons in Si(110) that have been addressed
in previous experimental
\cite{Backe_EtAl_2008,Backe_EtAl_2010,Backe_EtAl_2011,BackeLauth_2013} 
studies.
To this end, we have performed extensive calculations of the particles 
trajectories and the emitted radiation spectra formed in $L=50$ $\mu$m 
crystalline silicon.
Two examples of our calculations are presented in Fig.~\ref{Figures_01_02.fig}.

In both panels, the curves represent the spectral enhancement factor, i.e., the
spectral distribution $\d E/\d (\hbar\om) $ of the radiation emitted
in the crystalline medium normalized by that in the amorphous silicon.
The latter was calculated within the Bethe-Heitler approximation \cite{Tsai1974}.

For 270 MeV incident energy (left panel) the presented spectrum, which corresponds to the
aperture 0.24 mrad, was averaged over $N_0=3500$ simulated electron trajectories. 
At the crystal entrance, the initial velocity for each trajectory was parallel 
to the Si(110) plane and aligned along the  $[10,-10,1]$ crystallographic direction.
The error bars reflect the statistical uncertainties which correspond to the
probability $0.999$. 
 
For 855 MeV electrons (right panel) the spectra were computed for the aperture 
0.21 mrad and for three different 
directions of the initial velocity as indicated.
The numbers of simulated trajectories  are: $N_0=3120$ for the $[10,-10,1]$ direction,
  $N_0=1919$ for $[8,-8,1]$, and $N_0=1679$ for $[13,-13,1]$.
The goal of these calculations was to analyze the sensitivity of the 
channeling radiation to the initial (random) direction  along the (110) plane.
We can state that all spectra are indistinguishable within statistical errors.  

The theoretical results for the enhancement factors were compared to the 
experimental data \cite{BackeLauth_2013}.
The analysis of the theory-versus-experiment will be published elsewhere.
Here we just note that perfect correspondence, both in the peak position and
height,  was found for 270 MeV incident electron beam.
For 855 MeV electrons, theoretical curves reproduce the peak position but 
underestimate the intensity of the channeling radiation by approximately 25-30 per cent.
The analysis of the source of this discrepancy is currently under way.

\section{855 MeV Electrons and Positrons in Straight and Bent Si (111)}

The structure of a silicon single crystal implies that a Si(111) planar channel
contains two (111) planes separated by the distance $d_{\rm n} = 0.784$ \AA\ .
Total width of the channel is $d = 3.136$ \AA.

The presence of two planes of crystal atoms in a single channel 
leads to specific features of the channeling oscillations for both
negatively and positively charged projectiles. 
These features are absent in the case of channeling along (100) or/and (110)
planes.
  
\begin{figure} [h]
\centering
\includegraphics[scale=0.275,clip]{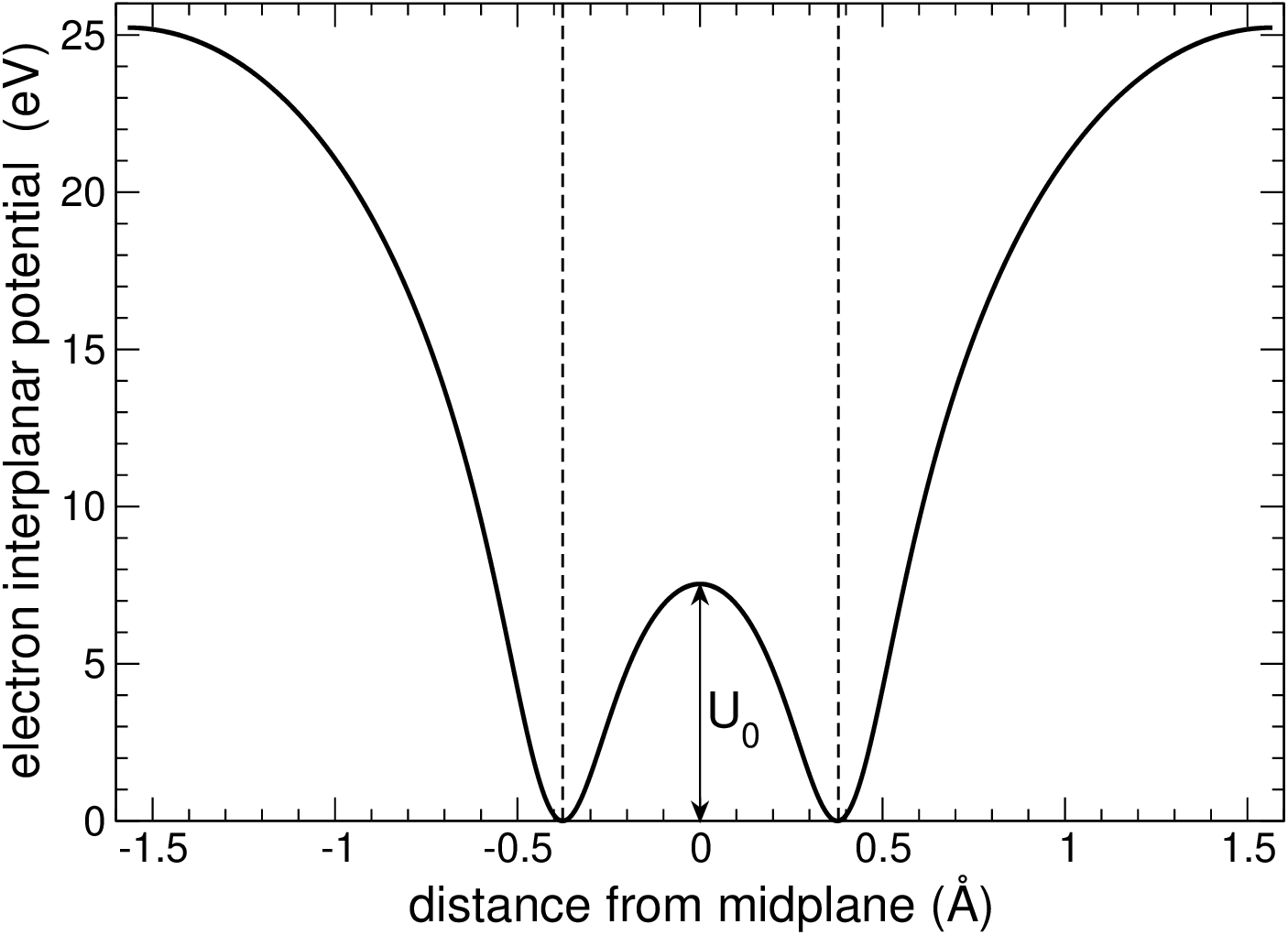}
\includegraphics[scale=0.275,clip]{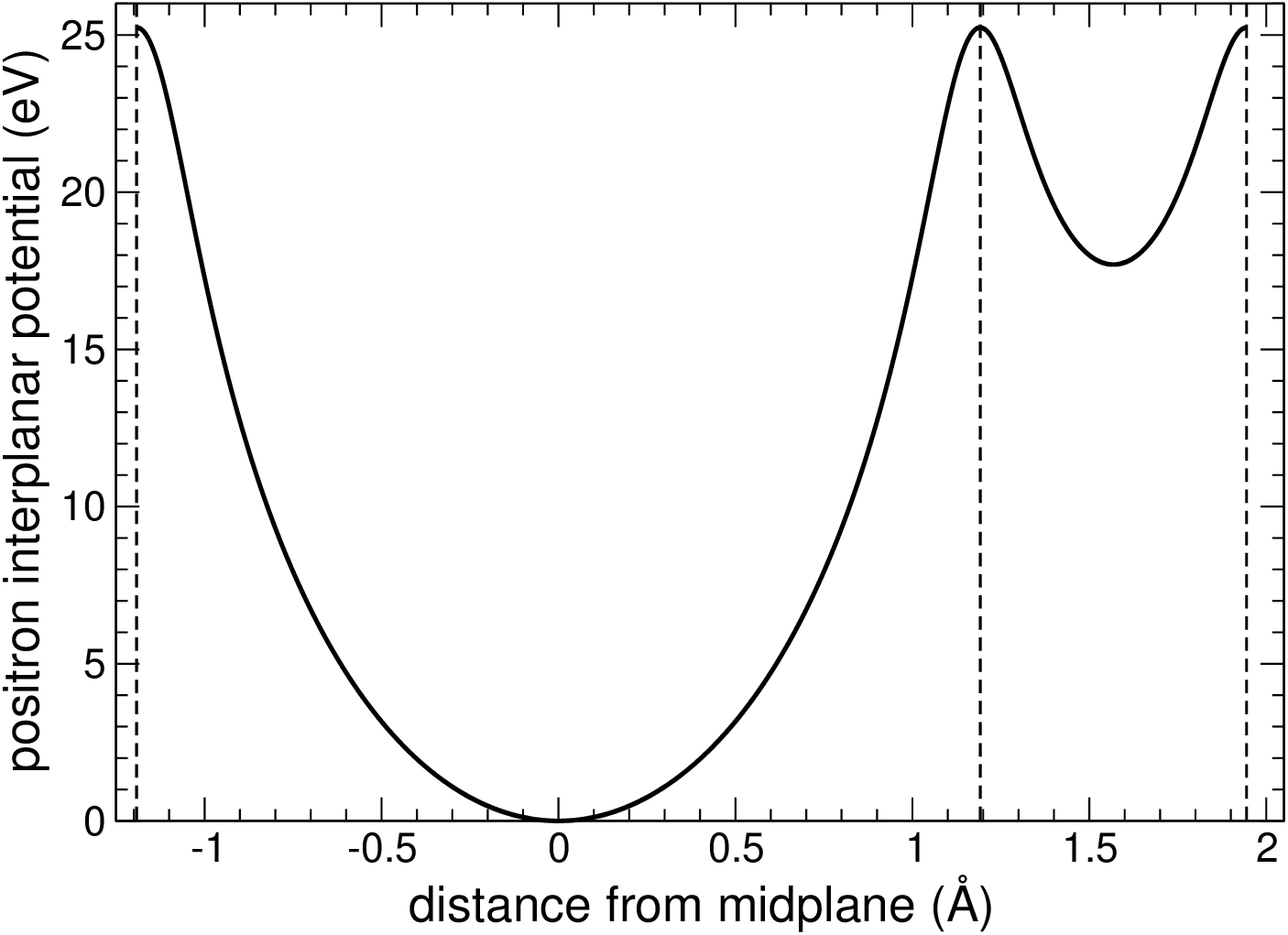}
\caption{
Electron (left panel) and positron (right panel) {\em continuous} Moli\`{e}re
interplanar potential for Si(111). 
Vertical dashed lines mark the positions of the atomic planes. 
The curves correspond to the crystal temperature 300 K.
}
\label{Figures_03_04.fig}
\end{figure}

To discuss qualitatively the channeling oscillations of an electron and a positron
we refer to Figure \ref{Figures_03_04.fig}, which presents the Si(111) interplanar 
potential $U$ calculated in the continuous approximation \cite{Lindhard}
with the use of the Moli\`{e}re atomic potential.
Let us stress, that the results of numerical simulations presented further 
in this section, were not based on the continuous approximation. 
The latter is used for illustrative considerations only.

In the case of electron channeling (left panel in Figure \ref{Figures_03_04.fig})
the interplanar potential has two wells symmetrically separated with 
respect to the midplane, where the potential has a local maximum $U_0$.
At the boarders of the channel, i.e. at the distances $\pm d/2$, the
potential has maxima $U_{\max}$ which exceed $U_0$.
As a result, if the transverse energy $\E_{\perp}$ of an electron 
satisfies the condition $\E_{\perp} < U_0$ then the channeling oscillations 
are restricted to one of the wells.
In the case $\E_{\perp} > U_0$ the particle oscillates with larger amplitude 
within $[-d/2,d/2]$.
The electron interplanar potential is strongly anharmonic, therefore, the period of 
oscillations depends on the amplitude.
On average, the large-amplitude oscillations are slower than 
the small-amplitude ones in the vicinity of the local minima.

The positron interplanar potential is presented on the right panel 
of Fig.~\ref{Figures_03_04.fig}. 
In this case, the potential also has two wells although strongly 
asymmetric. 
Both of the wells can be approximated by parabolic dependencies.
The frequency of channeling oscillations in the narrow (and shallow) well
is approximately 2 times larger than that in the wide well.   

Two types of channeling oscillations for electrons and positrons 
manifest themselves in the emission spectra calculated for small emission angle
and discussed below in the paper.

\begin{figure} [h]
\centering
\includegraphics[scale=0.32,clip]{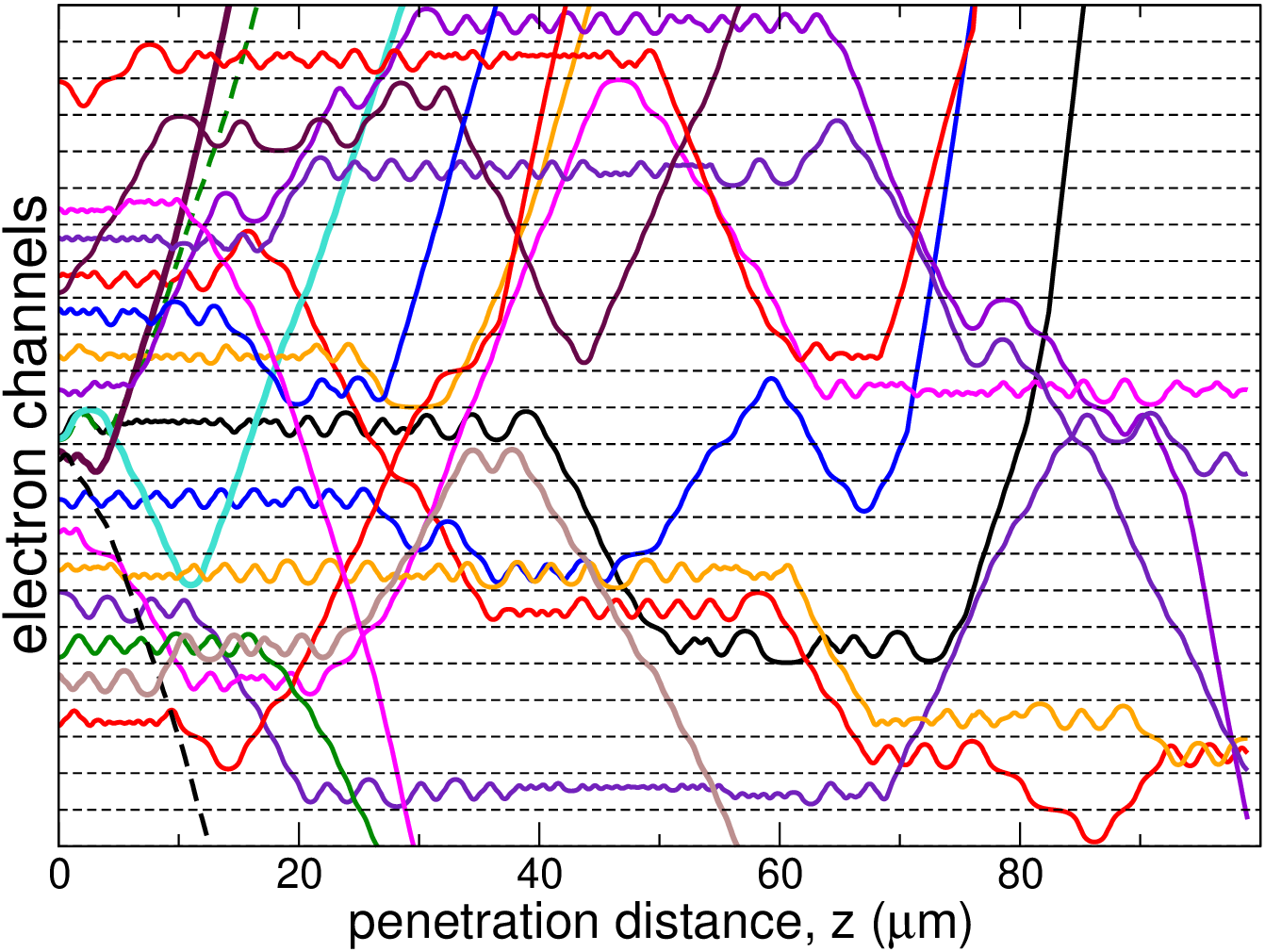}
\hspace*{0.5cm}
\includegraphics[scale=0.32,clip]{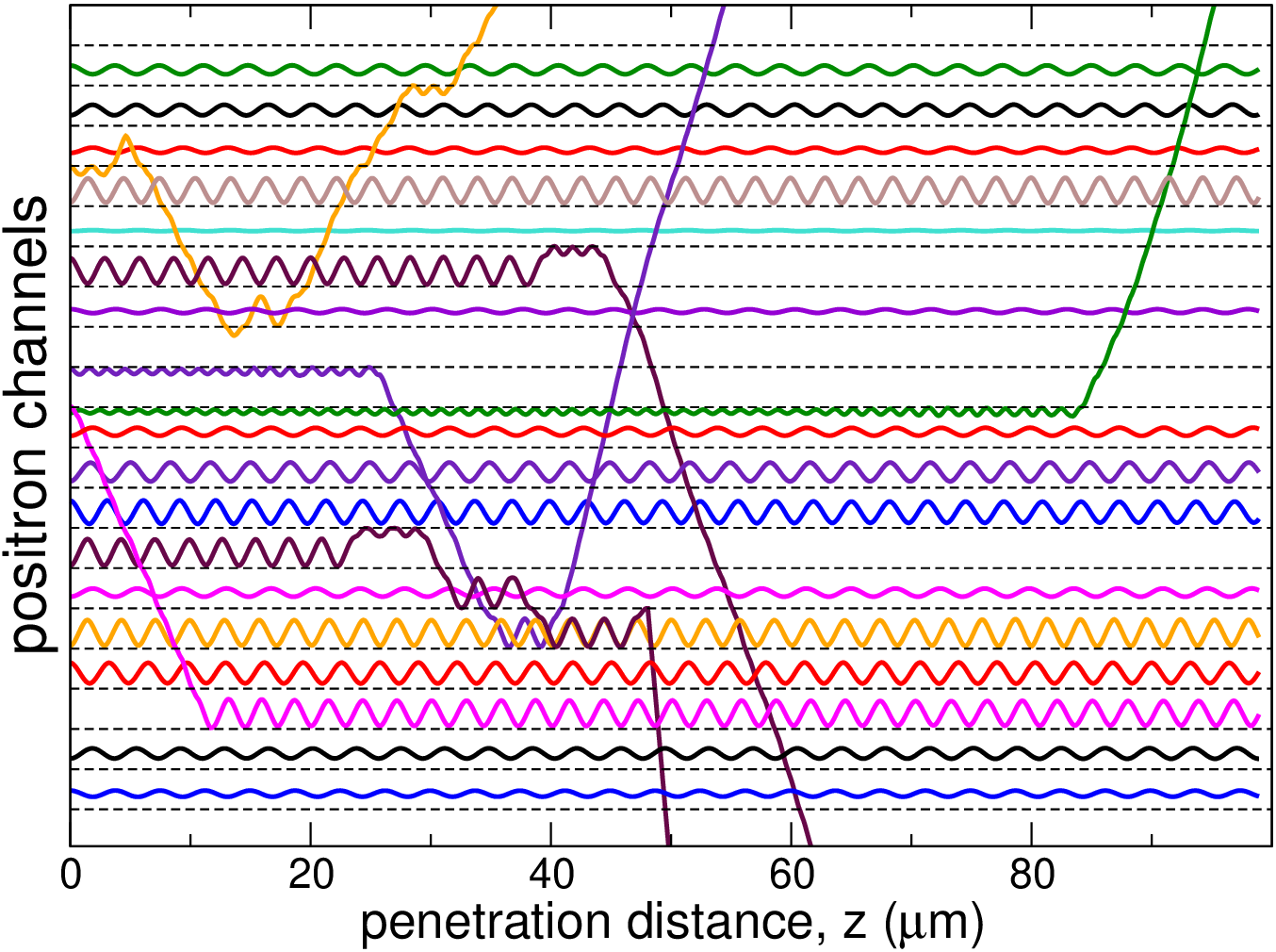}
\caption{
Channeling of 855 MeV electrons (left panel) and positrons (right panel)
in a $100$ $\mu$m thick silicon crystal.
The plots show randomly chosen trajectories of the particles initially collimated
along Si(111) crystallographic planes.
Horizontal dashed lines indicate the (111) planar channels  separated by the
distance $d=3.136$ \AA.}
\label{Figures_05_06.fig}
\end{figure}

The motion of 855 MeV electrons and positrons collimated at the entrance along 
Si(111) planes
is illustrated in Fig.~\ref{Figures_05_06.fig}
by sets of randomly chosen simulated trajectories.
The data refer to the straight crystal of the length $L=100$ $\mu$m. 

For positrons, noticeable are nearly harmonic oscillations.
Two types of oscillations, occurring in the wide part of the 
channel and in the narrow as well, are clearly seen in the presented trajectories.
Another feature of  positron channeling through a $L=100$ $\mu$m thick crystal
is a comparatively small number of the dechanneling events.
This is also not surprising if one compares the crystal size with the dechanneling
length $\Ld\approx 700$ $\mu$m for a 855 MeV positron in Si(111).
The latter value can be obtained using Eq. (1.50) from
\cite{BiryukovChesnokovKotovBook} with the correction for a light projectile
introduced in  \cite{Dechan01}.
Therefore, it is not surprising that most of the incident particles
traverse the crystal in the channeling mode.

Much less regular are the channeling oscillations of electrons, see
the left panel of Fig.~\ref{Figures_05_06.fig}.
The electron trajectories exhibit a broader variety of
features: channeling motion, over-barrier motion,
rechanneling process, rare events of hard collisions etc.
First, let us note that the dechanneling length of a 855 MeV electron in Si(111),
estimated with the help of Eq. (10.1) from \cite{Baier}, is $\Ld\approx 23$ $\mu$m.
Therefore, it is not surprising that the events of channeling through the 
whole crystal are quite rare.
On the other hand, the events of rechanneling, i.e., capture to the channeling 
mode of an over-barrier particle, are quite common for electrons.
Even the multiple rechanneling events are not rare.
This phenomenon has been already noted in the 
simulations of the electron channeling 
\cite{KKSG_simulation_straight,NewPaper_2013}
with a qualitative explanation provided \cite{KKSG_simulation_straight} 
of the difference in the
rechanneling rate for positively and negatively charged projectiles.
Also it is worth noting a visible anharmonicity in the channeling oscillations of
electrons which is a direct consequence of a strong deviation of the
the electron interplanar potential from a harmonic shape.
As a result, the period of the oscillations varies with the amplitude.
Similar to the positron channeling, two types of oscillations, related to the 
two wells structure of the interplanar potential (see Fig.~\ref{Figures_03_04.fig} left),
are clearly seen in the presented trajectories.

The simulations of the trajectories were performed for straight and bent Si(111)
channels.
In the former case, two crystal lengths along the incident beam direction
were considered:  
$L_1=50$ $\mu\mbox{m}$ and $L_2=100$ $\mu\mbox{m}$.
For the bent crystal (the uniform bending with the bending radius $R=33$ mm)
the simulations were carried out for $L_1=50$ $\mu$m in accordance with the 
experimental conditions at the Mainz Microtron facility \cite{BackeLauth_2013}.

The simulated trajectories were used to estimate the dechanneling length 
(in the case of the electron channeling) and to calculate spectral distribution
of the emitted radiation.

To quantify the electron dechanneling process we calculated two 
penetration lengths introduced in Ref. \cite{NewPaper_2013}. 
The first one, notated below as $L_{\rm p1}$ was found as a mean value
of the primary channeling segments, which started at the entrance and lasted till
the dechanneling point somewhere inside the crystal.
Generally speaking, this quantity is dependent on the angular distribution
of the particles at the entrance.
The $L_{\rm p1}$ values quoted below were obtained for a zero-emittance beam
collimated initially along the (111) planar direction.
The second penetration depth, $L_{\rm p2}$, is
defined as  a mean value of all channeling segments, including those which
appear due to the rechanneling.
In the rechanneling process an electron is captured into the channeling mode
having, statistically, an arbitrary value of the incident angle $\psi$
not greater than  Lindhard's critical angle.
Therefore, $L_{\rm p2}$ mimics the penetration depth of the beam with
a non-zero emittance $\approx \psi_{\rm L}$.

In addition to $L_{\rm p1}$ and $L_{\rm p2}$ we calculated 
the total channeling length $\Lch$, 
defined as an average length of all channeling 
segments per trajectory.

The results for $L_{\rm p1}$, $L_{\rm p2}$ and $\Lch$, 
together with the calculated values of the channel acceptance 
$\calA = \Nacc/N_0$
(where $N_0$ and $\Nacc$ are numbers of the incident and the accepted particles,
respectively),
are summarized in Table \ref{Si111.Table}.

\Table{\label{Si111.Table}
The penetration lengths $L_{p1}$, $L_{p2}$ and $L_{\rm chan}$ for  
$\E=855$ MeV electrons in straight and bent Si (111) crystals (the bending radius 
is 33 mm).
The crystal lengths are $L_1=50$ $\mu\mbox{m}$ and $L_2=100$ $\mu\mbox{m}$.
Also indicated are:
the number of simulated trajectories $N_0$,
the number of accepted particles $\Nacc$, 
and the acceptance $\calA = \Nacc/N_0$.
}
\br
Crystal   &$N_{0}$&$N_{\rm a}$&  ${\cal A}$ &  $L_{p1}$  &  $L_{p2}$  &  $L_{\rm chan}$  \\
          &       &           &             &  ($\mu$m)  &  ($\mu$m)  &  ($\mu$m)       \\
\br
straight $L_1$     & 3467   &  2560    &  0.74     &$18.37\pm0.82$&$15.48\pm0.52$&$24.72\pm0.83$ \\
straight $L_2$     & 2160   &  1632    &  0.76     &$19.22\pm1.24$&$16.38\pm0.61$&$38.71\pm1.76$ \\
straight $L_1$ \& $L_2$
          & 5627   &  4192    &  0.75     &$18.70\pm0.69$&$15.92\pm0.40$&\dash \\
\mr
bent      $L_1$     & 3603   &  2471    &  0.69     &$16.62\pm0.77$&$15.96\pm0.67$&$19.15\pm1.11$ \\
\br
\end{tabular}
 \end{indented}
 \end{table}

For a straight crystal, it is instructive to compare the 
obtained values $L_{p1}=18.70\pm0.69$ and 
$L_{p2}=15.92\pm0.40$ $\mu$m with the dechanneling lenghts for the initial beam,
$L_{\rm d 0}=13.57 \pm 0.12$ $\mu$m, and
for the rechanneled particles, $L_{\rm d}=13.69 \pm 0.07$ $\mu$m obtained 
in Ref. \cite{KKSG_simulation_straight}.
The calculations performed in the cite paper were based
on the peculiar model of the elastic scattering of an ultra-relativistic projectile
from the crystal constituents.
The model substitutes the atom with its ``snapshot'' image:
the atomic electrons are treated as point-like charges placed at fixed
positions around the nucleus.
The interaction of an ultra-relativistic projectile (e.g., an electron) with 
each atomic constituent is treated in terms of the classical Rutherford scattering.
In Ref. \cite{NewPaper_2013} it was demonstrated, that such
a ``snapshot'' model noticeably overestimates the mean scattering angle
in the process of elastic scattering in a single electron-atom collision.
The mean square angle for a single scattering is a very important quantity
in the multiple-scattering region,
where there is a large succession of small-angle deflections symmetrically
distributed about the incident direction.
It was noted in Ref. \cite{NewPaper_2013} that the ``snapshot'' approximation
underestimates the dechanneling length of 855 MeV electrons in straight Si (110) 
by approximately 30 per cent.
Similar to this, the Si(111) data from \cite{KKSG_simulation_straight},
undervalues the dechanneling length presented in Table \ref{Si111.Table}:
$L_{\rm d 0}$ is less than $L_{p1}$ by $37 \pm 5$ \% 
whereas $L_{\rm d}$ is $17  \pm 3$\% smaller than $L_{p2}$.

\textcolor{black}{Let us note that the obtained length $L_{p1}=18.70\pm0.69$ $\mu$m
is in agreemeent with the value $18.8$ $\mu$m evaluated recently in 
Ref. \cite{BogdanovDabagov2012} from the solution of the Fokker-Plank equation.}

\begin{figure} [h]
\centering
\includegraphics[scale=0.32,clip]{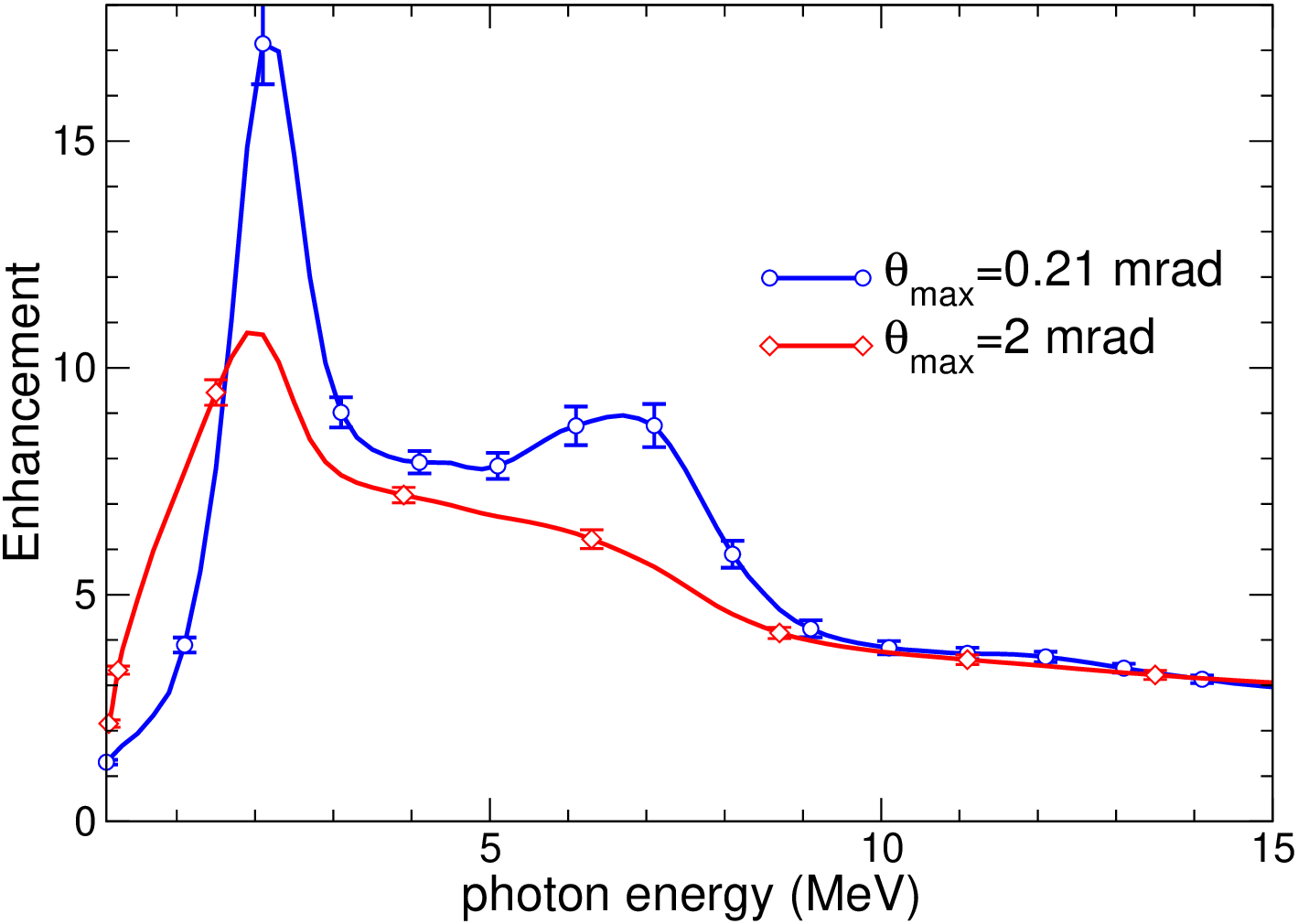}
\includegraphics[scale=0.32,clip]{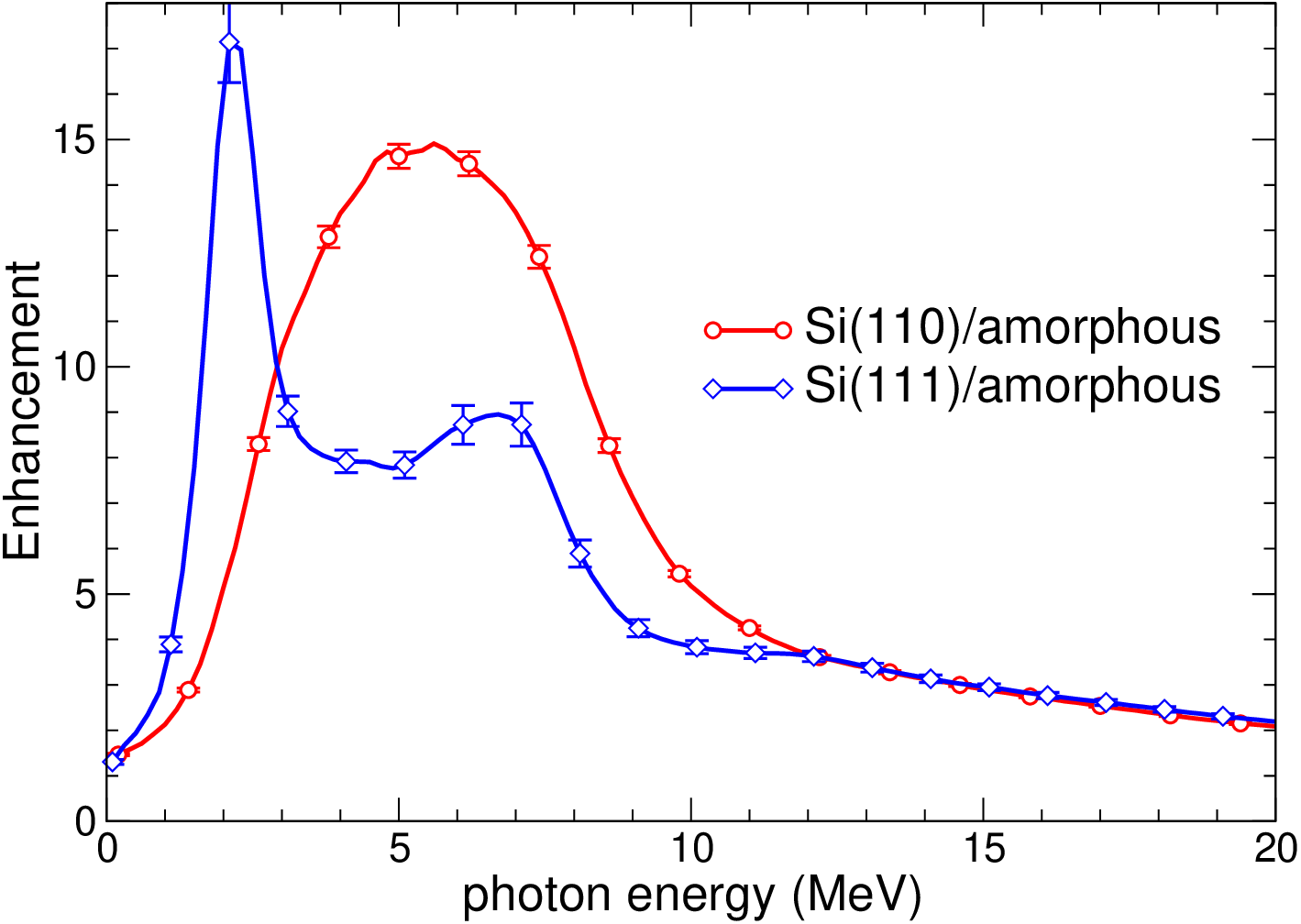}
\caption{
{\em Left panel.} 
Enhancement factors,
calculated for two indicated apertures $\theta_{\max}$, 
for 855 MeV {\em electron} channeling in 
$L=50$ $\mu$m straight Si crystal along (111) planes
\\
{\em Right panel.} 
Enhancement factor for 855 MeV {\em electrons} channeled in 
$L=50$ $\mu$m straight silicon crystal along (110) and (111) planes.
The data refer to $\theta_{\max}=0.21$ mrad.
}
\label{Figures_08_09.fig}
\end{figure}

\begin{figure} [h]
\centering
\includegraphics[scale=0.32,clip]{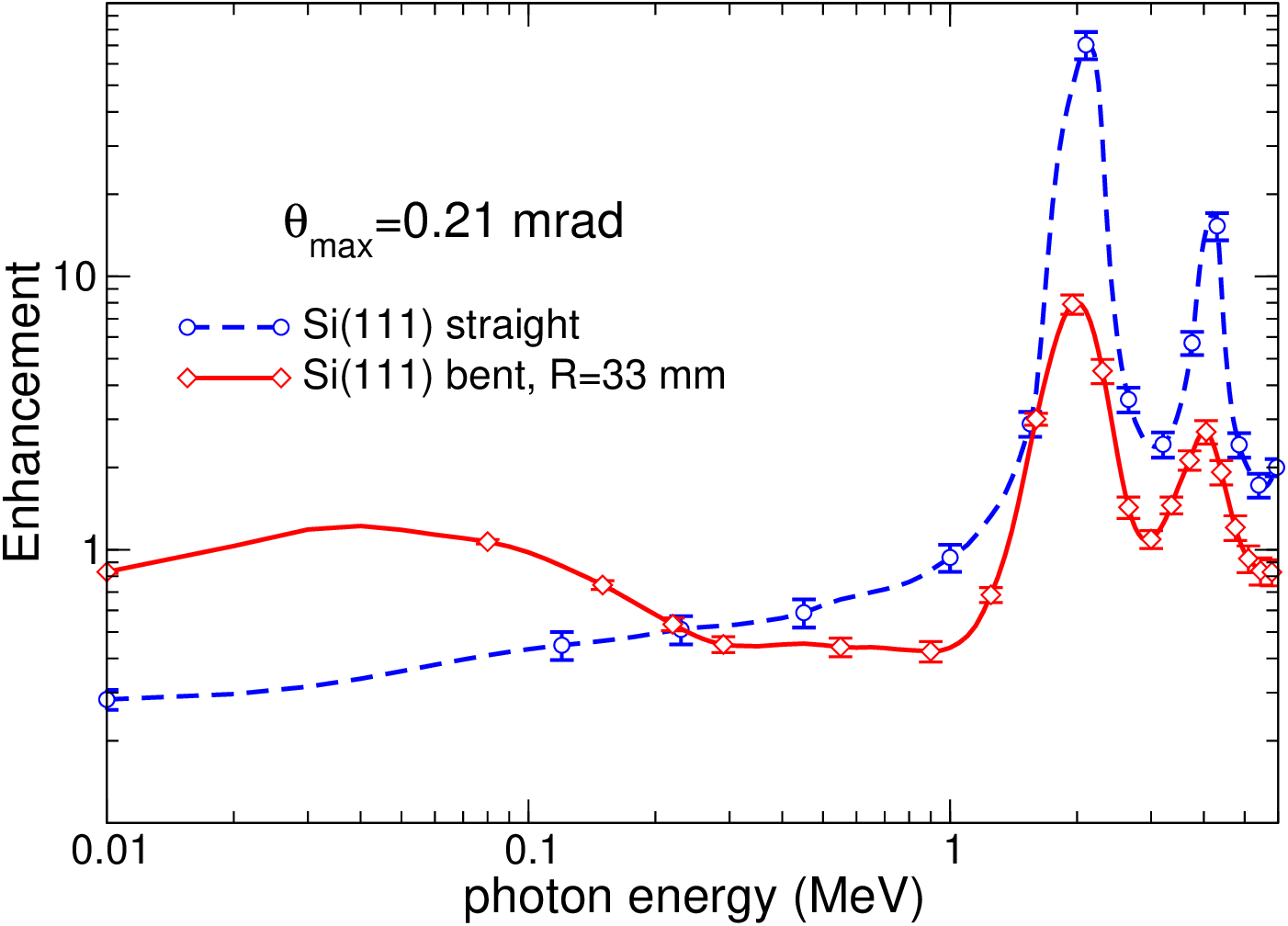}
\includegraphics[scale=0.32,clip]{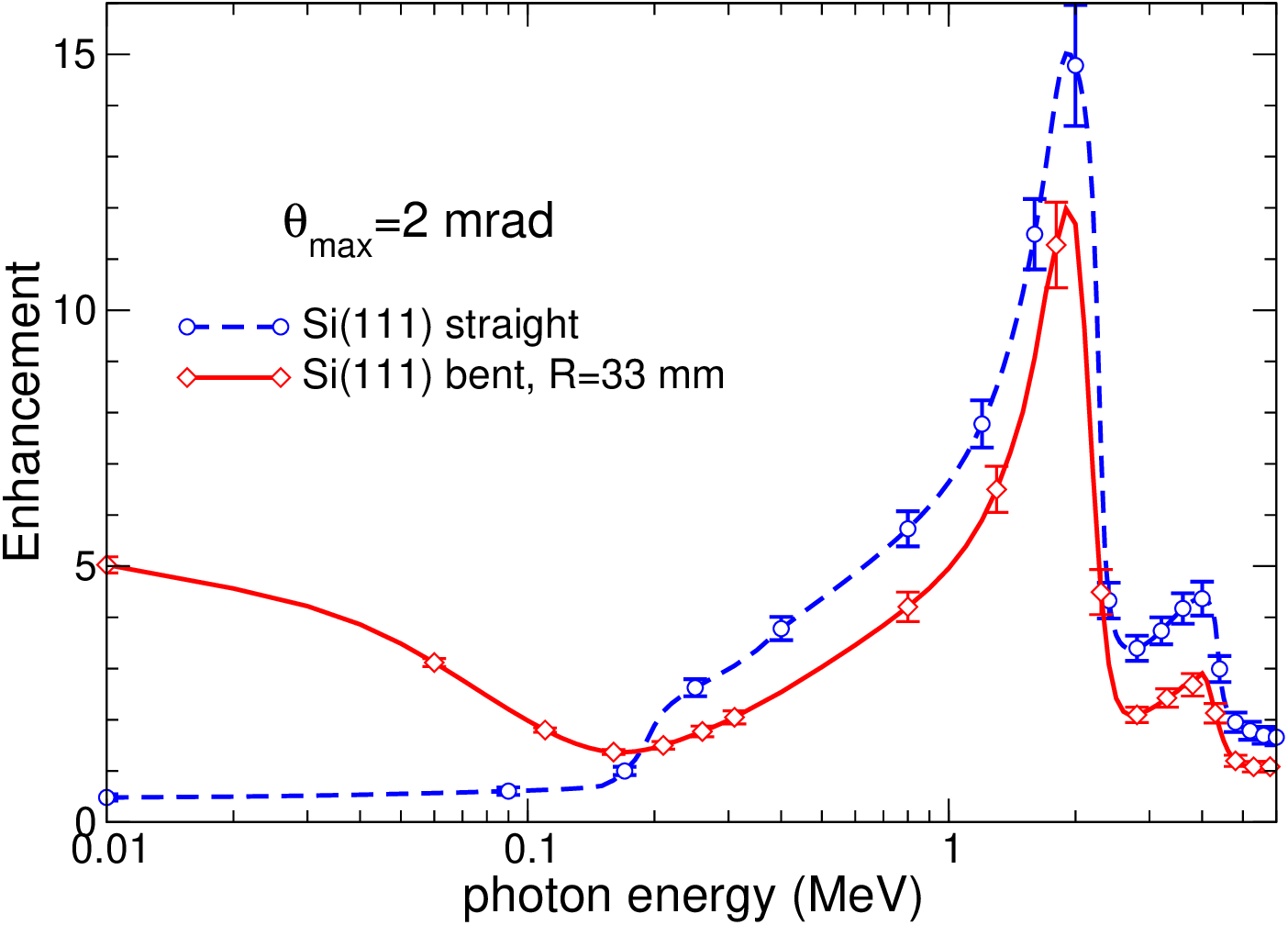}
\caption{
Enhancement factor calculated for two apertures $\theta_{\max}$
for 855 MeV {\em positrons} channeled along
straight (dashed line) and bent (solid line) Si (111) plane.
The bending radius $R=33$ mm, the crystal length $L=50$ $\mu$m.
Note the double log scale in the left graph. 
}
\label{Figures_10_11.fig}
\end{figure}

The simulated trajectories were used to compute
spectral distribution of the emitted radiation following the formalism and 
algorithm described in detail in Ref.~\cite{NewPaper_2013}. 
The results are presented in figures 
\ref{Figures_08_09.fig} -- \ref{Figures_10_11.fig}.
The calculated spectral intensities are normalized to the Bethe-Heitler 
values (see, for example, Ref. \cite{Tsai1974}) and, thus, are
plotted as the enhancement factors over the bremsstrahlung spectrum in
amorphous silicon.
Statistical uncertainties due to the finite number ($\approx 3000 \dots 4000$)
of the analyzed trajectories are indicated by the error bars.
The calculations were performed for two detector apertures: 
$\theta_{\max}=0.21$ and $2$ mrad. 
The first value, which is close to the aperture used in the experiments with
the 855 MeV electron beam \cite{Backe_EtAl_2008,Backe_EtAl_2011,BackeLauth_2013}, 
is much smaller than the natural emission angle $\gamma^{-1} \approx 0.6$ mrad.
Therefore, the corresponding spectra refer to nearly forward emission.
On the contrary, the second angle greatly exceeds $\gamma^{-1}$, so that 
the cone $\theta_{\max}$ collects nearly all emitted 
radiation.\footnote{\textcolor{black}{The 
intensity of channeling radiation of 855 MeV electrons in Si(111) crystals of
various thickness was calculated in \cite{BogdanovDabagov2012} for large apertures.
In the cited paper the channeling process was modeled on the basis of the Fokker-
Plank equation.}}   

\begin{figure} [h]
\centering
\includegraphics[scale=0.35,clip]{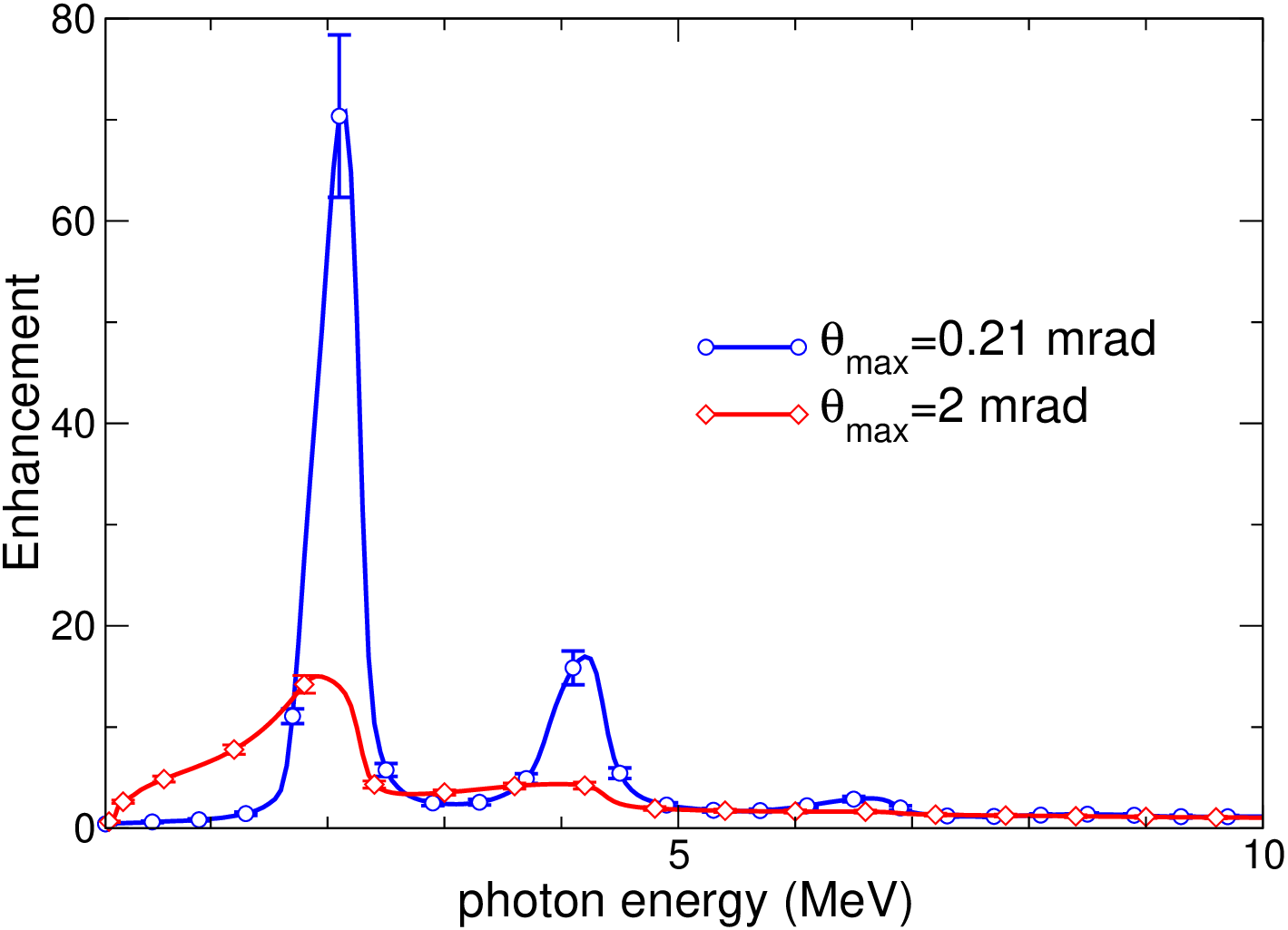}
\caption{
Enhancement factor for 855 MeV {\em positrons} channeled in 
$L=50$ $\mu$m crystalline target along Si (111) planes
calculated for two values of the  maximum emission angle 
as indicated.
}
\label{Figure_07.fig}
\end{figure}

Figures \ref{Figure_07.fig} and \ref{Figures_08_09.fig}
present the enhancement of radiation in straight silicon crystals.

First, we note that for both electrons and positrons the intensity of radiation in
the oriented crystal greatly exceeds (by more than an order of magnitude)
the bremsstrahlung background.
The enhancement comes from the particles moving along quasi-periodic 
channeling trajectories, which bear close resemblance with the undulating motion.
As a result, constructive interference of the waves emitted from different
but similar parts of the trajectory are added coherently.
For each value of the emission angle $\theta$ the coherence
is most pronounced for the radiation into harmonics,
which frequencies can be estimated as
(see, e.g., \cite{Baier}):
\begin{eqnarray}
\om_{n}
=
{2\gamma^2\, \Om_{\rm ch}
\over
1 + \gamma^2 \theta^2 + K_{\rm ch}^2/2}
\, n ,
\quad
n=1,2,3,\dots \,,
\label{Section2.05_2:eq.01}
\end{eqnarray}
where $\Om_{\rm ch}$ is the frequency of channeling oscillations
and $K_{\rm ch}^2 = 2\gamma^2 \left\langle v_{\perp}^2\right\rangle/c^2$
is the mean square of the undulator parameter related to them.
Within the framework of continuous potential approximation, these quantities
are dependent on the magnitude of the transverse energy which, in turn,
determines the amplitude of oscillations.

Different character of channeling by positrons and electrons results
in differences in the spectra of the channeling radiation.

The nearly perfect sine-like channeling trajectories of positrons
lead to the emission spectrum close to that of the undulator radiation with
$K^2 \ll 1$.
Two peaks in the positron spectrum, see  Fig.~\ref{Figure_07.fig}, 
is due to two types of channeling oscillations mentioned above.
The peak at $\approx 2$ MeV is due to the emission in the fundamental 
harmonic ($n=1$) from the trajectories corresponding to the channeling motion
in the wide well of the positron Si(111) channel (see Fig. \ref{Figures_03_04.fig} right).
It is more pronounced for the smaller aperture, since in this case 
a strong inequality $(\gamma \theta)^2\ll 1$, valid for all angles
$\theta \leq \theta_{\max}$, ensures the independence of $\om_{1}$ 
on the emission angle.   
The second, less accented peak, corresponds to the emission in
the first harmonic due to the channeling motion in the narrow part of the channel.
In this case, the amplitudes are smaller (this result in the decrease of the intensity)
but the channeling frequencies are higher leading to the higher value of $\om_{1}$.
For the larger aperture, a big part of the energy is radiated into the cone
$\gamma^{-1} < \theta < \theta_{\max}$. 
For these relatively large emission angles the first harmonic energy decreases with
$\theta$.
As a result, the peaks become broader and less intensive.

Due to strong anharmonicity of the electron channeling oscillations, 
the peaks in the spectrum of channeling radiation are less pronounced even for 
the small aperture, see Fig.~\ref{Figures_08_09.fig} left.
For the large aperture  $\theta_{\max} =2$ mrad the second peak 
is completely smeared out.
The right panel in the figure illustrates the differences in the emission spectra 
(for the small aperture) for electron channeling in Si (111) and Si (110) channel
(the calculations of the latter were performed in \cite{NewPaper_2013}).
The Si(110) channel can be modeled as a single-well interplanar potential which leads
to a single peak in the emission spectrum.   

\begin{figure} [h]
\centering
\includegraphics[scale=0.32,clip]{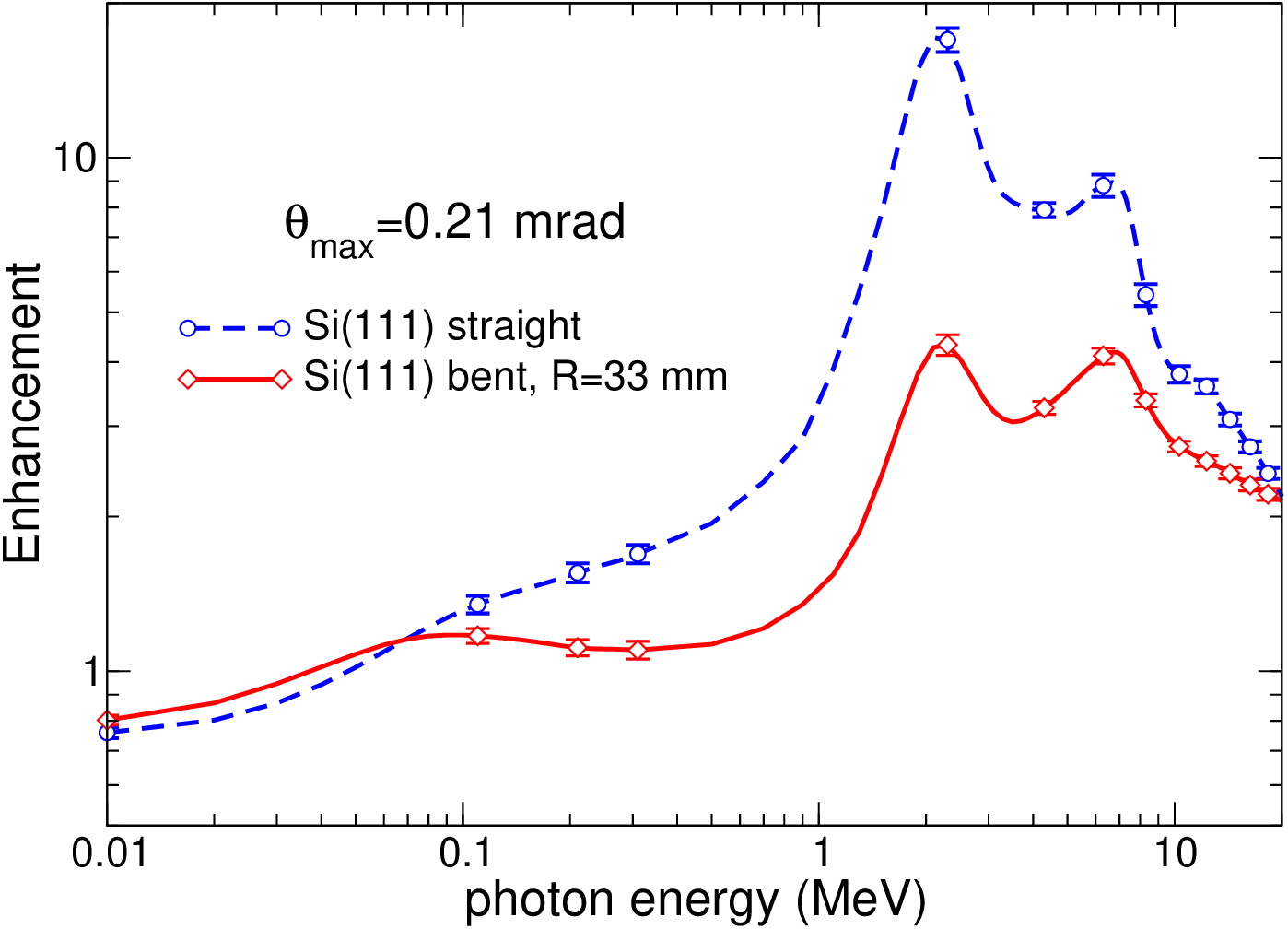}
\includegraphics[scale=0.32,clip]{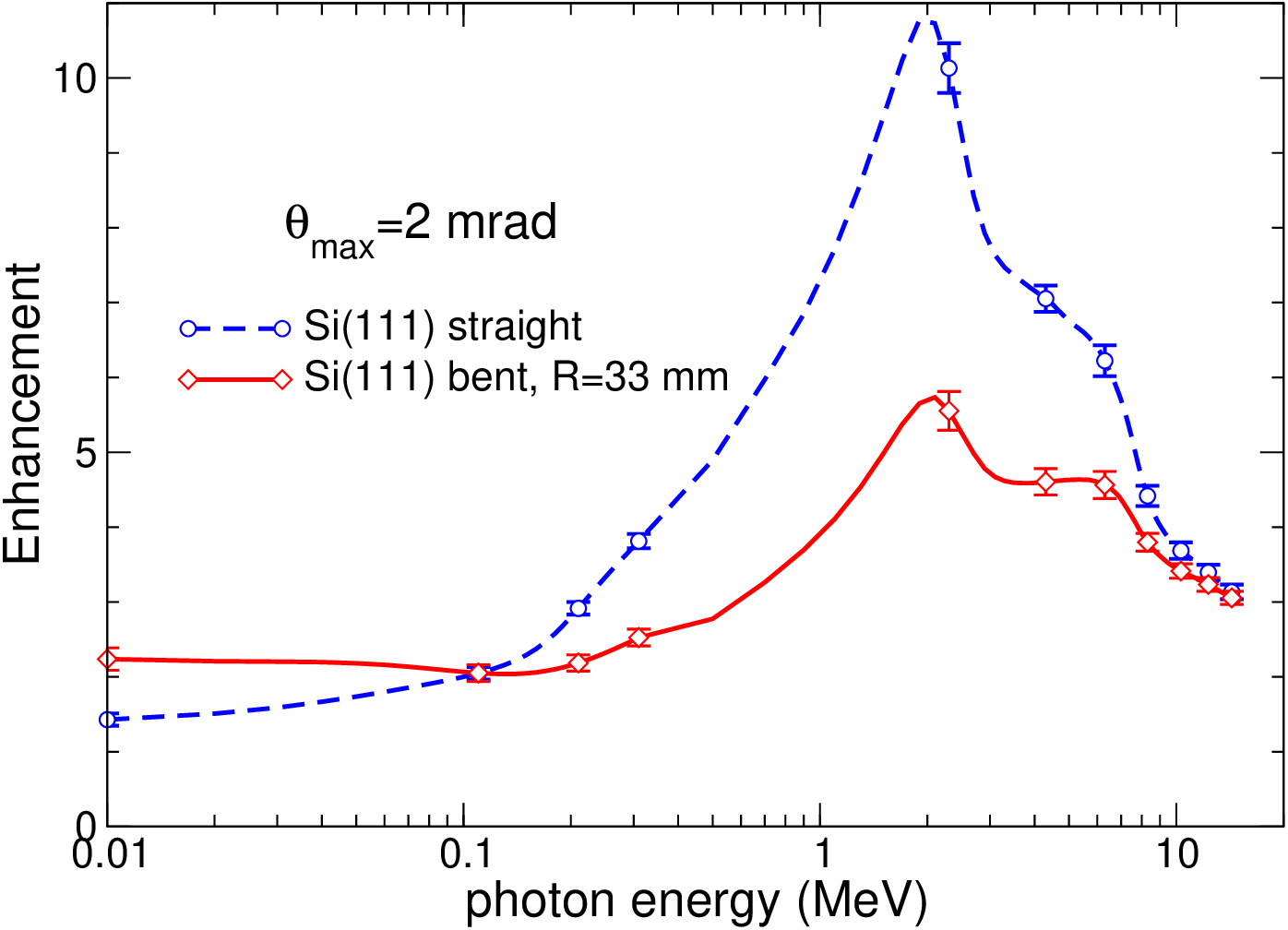}
\caption{
Same as in Fig.~\ref{Figures_10_11.fig} but 
for  855 MeV {\em electrons}.
}
\label{Figures_12_13.fig}
\end{figure}

Enhancement of radiation emitted by 855 MeV positrons and electrons in 
uniformly bent Si (111) channels are presented in 
Figs.~\ref{Figures_10_11.fig} and \ref{Figures_12_13.fig}.
The data correspond to the crystal length $L=50$ $\mu$m and the bending radius 
$R=33$ cm.
In a bent crystal, the channeling condition implies the centrifugal force
$F_{\rm cf}\approx \E/R$ to be smaller than the maximum interplanar force
$U_{\max}^{\prime}$ \cite{Tsyganov1976}.
In our case, the value of the bending parameter 
$C=F_{\rm cf}/U_{\max}^{\prime}$ \cite{ChannelingBook2013} 
is much less than one, $C \approx 0.045$, so that the channeling condition 
is well-fulfilled.
This is reflected, in particular, by small deviation of the values of the
penetration lengths $L_{\rm p1}$, $L_{\rm p2}$ and $\Lch$ calculated for an 
electron channeled in the bent Si(111) from those for the straight channel, 
see Table \ref{Si111.Table}.

Let us note two features of the emission spectra formed in bent silicon crystal.
First, the bending gives rise to the synchrotron radiation, since the channeled particle
experiences the circular motion in addition to the channeling oscillations.
This leads to the increase of the intensity in the photon energy range $\lesssim 10^2$ keV. 
For these energies the excess of the radiation yield from the bent channel over 
that from the straight channel is clearly seen in the figures 
(compare the solid and the dashed curves).
Second, it is seen that the decrease in the intensity of the channeling radiation 
in the bent channel is more pronounced for the small aperture, $\theta_{\max}=0.21$ 
mrad than for the large one equal to $2$ mrad. 
This feature can be understood by comparing $\theta_{\max}$ to the bending angle
$\Theta=L/R \approx 1.5 \mbox{mrad} \gg \gamma^{-1}$. 
Since the emission angle is defined with respect to the direction of the incident beam,
the emission within the cone $\theta_{\max}=0.21 \ \mbox{mrad} \ll \Theta$ occurs, 
effectively, only from the initial parts of the channeling trajectories.
In the opposite limit, $\theta_{\max} > \Theta$, the radiation from nearly all the 
trajectory is detected.

\section{Channeling and Radiation in Crystalline Undulators}

In this section we present some of the results obtained recently with the 
help of the newly developed code \cite{NewPaper_2013}
for the channeling phenomenon and radiation emitted
in a crystalline undulator (CU).
In CU, in addition to the channeling radiation, 
the undulator-type radiation appears   
due to the undulating motion of channeling particles which follow the periodic 
bending of crystallographic planes \cite{KSG1998,KSG_review_1999,ChannelingBook2013}. 
Two types of harmonic periodic bending of the 
channel centerline, which correspond to the sine and to the cosine profiles,
can be considered:
\begin{eqnarray}
y(z) = a\sin\left(2\pi z/\lamu\right), 
\qquad
y(z) = a\cos\left(2\pi z/\lamu\right)\,.
\label{profile}
\end{eqnarray}
Here, the coordinate $z$ is measured along the straight channel, the 
$y$ axis is perpendicular to the straight plane. 
The quantities $a$ and $\lamu$ are the bending amplitude and period and they
satisfy the relation $d < a \ll \lamu$ where $d$ is the interplanar distance
(see the cited papers for more details on the description of the CU concept).
    
We have performed simulation of the trajectories, the quantitative 
analysis of the channeling motion and computation of the spectral intensities 
of the radiation formed  by ultra-relativistic
{\em electrons and positrons} within the energy range 195 \dots 855 MeV in 
the CU with the parameters used in the experiments at  
at the Mainz Microtron (Germany) facility
\cite{Backe_EtAl_2010,Backe_EtAl_2011}.
The 4-periods CUs were manufactured in Aarhus University (Denmark) 
using the molecular beam
epitaxy technology to produce strained-layer Si$_{1-x}$Ge$_{x}$ 
superlattices with varying germanium content as described in 
\cite{MikkelsenUggerhoj2000,Darmstadt01}.

The following values of the CU parameters were used in the calculations:
\begin{itemize}
\item
Channeling plane: Si(110) (the interplanar distance $d=1.92$ \AA)

\item
Crystal length: $L=39.6$ $\mu$m

\item
Bending period: $\lamu=9.9$ $\mu$m

\item
Bending amplitude: $a=3\dots 5$ \AA

\end{itemize}

\begin{figure} [h]
\centering
\includegraphics[scale=0.32,clip]{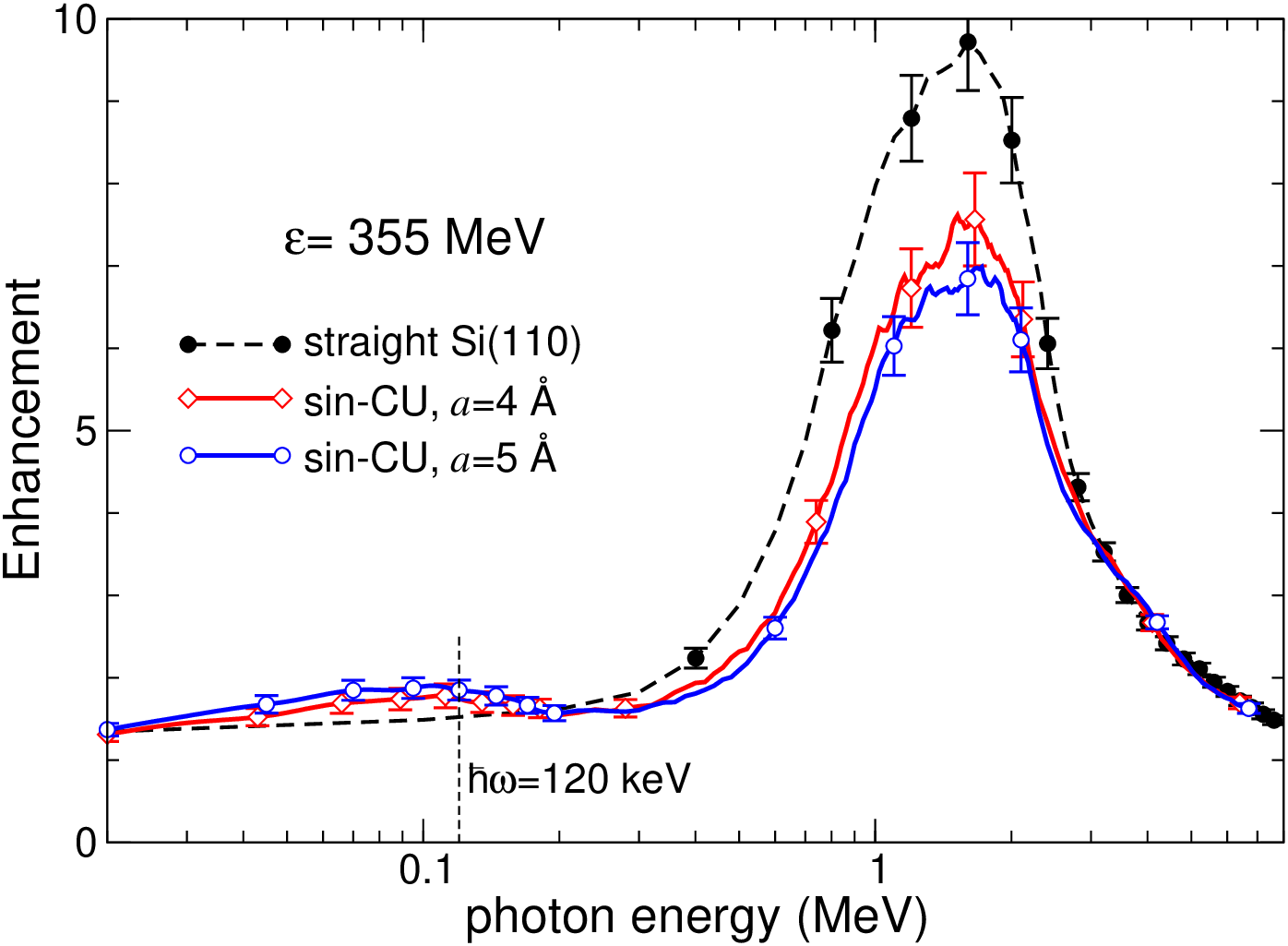}
\hspace*{0.3cm}
\includegraphics[scale=0.32,clip]{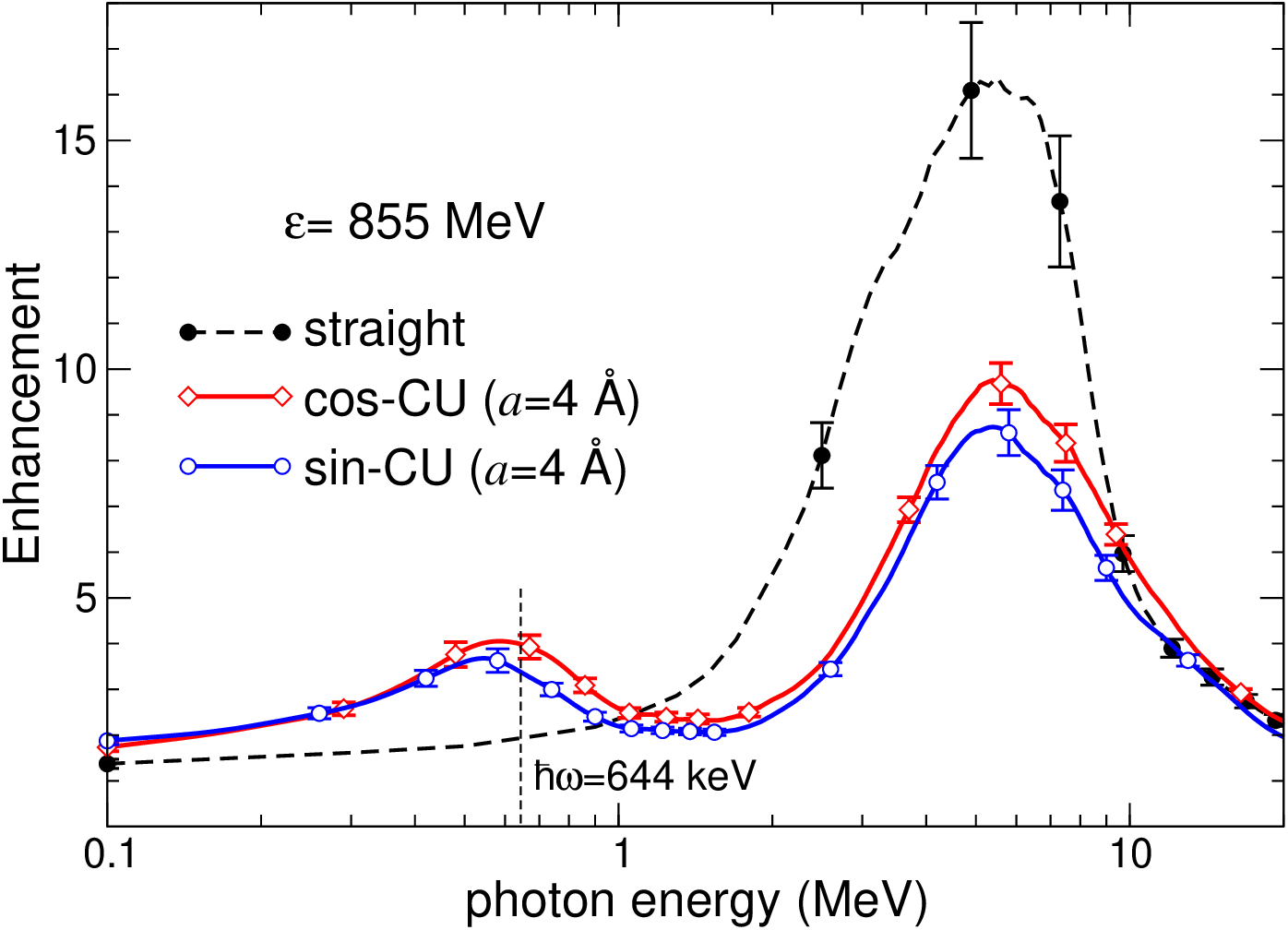}
\caption{
Enhancement of radiation emitted in straight Si (110) (dashed lines)
and in CU (solid lines) over the Bethe-Heitler spectrum.
\underline{Left panel:} the data refer to $\E=355$ MeV electrons.
Two solid lines present the dependencies for the sine-like CUs of the 
indicated amplitudes.
\underline{Righ panel:} the data refer to $\E=855$ MeV electrons.
Two solid lines stand for the spectra emitted from the sine- and cosine-like
CUs of the same amplitude.
The maximum emission angle is $\theta_{\max}=0.21$ mrad for both panels.
}
\label{Figures_14_15.fig}
\end{figure}

In full the results of our calculations will be published elsewhere.
Here we announce only few results for 355 and 855 MeV electrons which are 
presented in Fig.~\ref{Figures_14_15.fig}.
The left panel compares the spectral enhancement of radiation emitted by 
355 MeV electrons  channeled in straight Si (110) 
and in the sine-shaped CU with two bending amplitudes as indicated.
All three dependencies exhibit powerful maximum at about $1.7$ MeV which
corresponds to the channeling radiation. 
In the case of the undulating crystals the maxima are lower is due to the 
decrease in the allowed amplitude of the channeling oscillations in periodically
bent channel in comparison with the straight one \cite{Dechan01}.
However, the CU undulator radiation manifests itself as a hump in the photon
energy range 40\dots 100 keV (the vertical dashed line marks the 
first harmonic of the radiation in the forward direction).
Hence, our simulations indicate that it is possible to observe the CU radiation
even for comparatively low energies of the electron beam.

With the increase of the electron beam energy the CU radiation peak becomes more 
accented, as it is illustrated by the right panel. 
Here, the maximum at about 600 keV is seen for both sine- and cosine-shaped       
CUs.

\section{\textcolor{black}{Conclusion}}

\textcolor{black}{
Using the newly developed code \cite{NewPaper_2013}, 
which was implemented as a module in the MBN Explorer package 
\cite{MBN_ExplorerPaper},
we have performed the Monte Carlo simulations of trajectories of  
ultra-relativistic electrons and positrons in oriented straight, bent and
periodically bent single Si crystals. }

\textcolor{black}{
The simulated trajectories were used as the input data for numerical analysis 
of the intensity of the emitted radiation.
In the case of straight crystals the reduces to the channeling radiation emitted
atop the incoherent bremsstrahlung background. 
In a bent channel the spectrum is enriched by the synchrotron radiation due to the
circular motion of the projectile along the bent centerline.
In a periodically bent crystal, in addition to the channeling radiation, the
undulator-type radiation appears due to the periodicity of the bending.
In this paper, for the first time the characteristics of the crystalline undulator 
radiation were computed on the basis of the Monte Carlo simulation of electron
and positron trajectories.}
 
\textcolor{black}{
The calculation of the spectra as well as the numerical analysis of channeling 
conditions and properties (acceptance, dechanneling length) 
have been carried out in a broad range of the beam energies, 195 \dots 855 MeV. 
The obtained and presented results are of interest in connection with  
the ongoing experiments with electron beams at Mainz Microtron \cite{Backe_EtAl_2011}
and with possible experiments with the positron beam \cite{Backe_EtAl_2011a}.
}

\ack
We are grateful to Hartmut Backe and Werner Lauth for fruitful and
stimulating discussions. 
The work was supported by the European Commission CUTE 
(the CUTE-IRSES project, grant GA-2010-269131). 
The possibility to perform complex computer 
simulations at the Frankfurt Center for Scientific Computing is 
gratefully acknowledged.

\section*{References}

\end{document}